# A study of GenAI usage by Design Students
## Analysis of Survey Results and Journals of AI practices at the Politecnico di Milano in 2025/2026

Stefana Broadbent
Department of Design Politecnico di Milano
stefanamaja.broadbent@polimi.it

The results of a survey on the use of GenAI by the design students of the Politecnico di Milano raises major questions around the role of AI in the Design Process. A domain specific set of questions alongside the more general purpose probes about GenAI usage, delivers insights into the particular practices that are emerging in Design. The very high frequency of use of GenAI tools is concentrated in the initial stages of projects and does not affect the perception of project ownership or creativity. An analysis of AI journals kept by a class during research assignments confirms the range of AI supported activities but also the limited trust in GenAI leading to systematic individual and collective verification and augmentation of outputs. Together, the findings suggest that design students are reflectively experimenting with how GenAI tools can contribute to their creative design process.

### 1. Introduction

The wide adoption by university students of Generative AI tools like ChatGPT, Claude, Perplexity has been reported by numerous studies on the use of AI in higher education. As shown in the bibliometric review by Dai et al. (2026) [1] there have been more than 2700 studies analysing the potential and risks of GenAI in higher education (HE). Research spans HE institutions across the world and covers four key thematic clusters in GenAI: (1) Adoption, perception, and behavioural engagement with GenAI in Higher Education, (2) GenAI in research practices and professional/ applied HE areas, (3) Technical evaluation and academic utility of GenAI systems, and (4) Ethics, integrity, and governance of GenAI in Higher Education.

The extensive use of GenerativeGenAItools in education is confirmed by the research on vast dataset of conversations with chatbots. Anthropic's Economic Index of 2026 [2] which analysed more than 1 Million conversations over 3 months, reports that 7% of all the interactions are for academic purposes, by both students and lecturers. Coursework accounts for more than 12% of conversations. This is not counting the huge number of queries to do with coding and maths which account for more than 35% of the use cases and could include activities related to research and education.

Individual universities and departments have also been running local and comparative surveys on their students, confirming the wide adoption of GenAI for a range of educational activities. Generative AI (GenAI) from the UCLA Class of 2025 Senior Survey [3] for instance, included responses from 6,639 out of 9,337 seniors awarded degrees in the 2024-2025 academic year. The survey included questions on students' overall frequency of use of GenAI tools to support academic coursework, frequency of use for specific tasks related to particular degrees, and perspectives on the role of GenAI in education, including potential concerns. The majority of respondents (73%) affirmed they were using GenAI tools to support their coursework. The most frequently reported applications for these tools were to research a topic (60% of students reported this use), to brainstorm for writing projects or presentations (57%), and to study for exams (57%). Less common tasks included using tools to draft responses to written assignments (31% reported this use), generate code (30%), and translate languages for an assignment (25%).

The study by Chung et al. (2026) [4] used an anonymous, large-scale, cross-sectional online survey administered in late 2024 across all departments of four Australian universities located in Victoria, New South Wales and Queensland. The survey design went through iterative development, using focus-groups with students and teaching staff to validate the questions.
Survey items covered:

- students’ knowledge of and access to GenAI tools
- patterns of use and perceived usefulness
- attitudes and trust, including perceived risks and benefits
- views on GenAI in their studies, in educators’ practice, and in future professional life

The answers from over 8000 anonymous respondents, showed that over 80% of students had used GenAI for study-related tasks, with nearly half of them using it regularly, primarily for editing, summarising, and idea generation. Usefulness, rather than blind trust or rule compliance, appears to drive their engagement. Students adopted GenAI pragmatically, with low trust in its factual accuracy, but confidence in their own ability to manage it as a support tool for academic work.

More recently, the HEPI Student Generative Artificial Intelligence Survey 2026 [5] in the UK at its third edition, shows that 95% of students report using GenAI in some way and 94% say they use generative AI to help with assessed work. The proportion of students directly including AI-generated text in assessed work has risen to 12%, up from 8% in 2025 and 3% in 2024. Almost half (49%) of the respondents believe that GenAI has improved their student experience, particularly by saving time, improving understanding and providing instant support.

Finally a significant study of a random selection of all messages sent to ChatGPT on consumer plans between May 2024 and June 2025 (Chatterji et al. 2025) [6] categorised the typology of requests. While not directly designed to analyse behaviours in higher education, the analysis sheds light on the role of GenAI in learning processes. Nearly 80% of all ChatGPT usage falls into three categories, which the authors called the Practical Guidance, Seeking Information, and Writing. Practical Guidance includes activities like tutoring and teaching, advice on how-to perform some tasks. Seeking Information includes searching for information about people, current events, products, and recipes, and appears to be very similar to web search. Writing includes automated writing of emails, documents and other communications, but also editing, critiquing, summarising, and translating text provided by the user. About two-thirds of all Writing messages ask ChatGPT to modify user text (editing, critiquing, translating, etc.) rather than creating new text from scratch. About 10% of all prompts are requests for explanations or teaching, suggesting that education in its broader sense, is a key use case for ChatGPT.

All the studies mentioned above, address general purpose GenAI usage in education, whereas the present study addresses the specific practices emerging in Design Education. Design involves a process that has multiple and very diverse phases, ranging from secondary research to understand the context (social, economic, regulatory, physical, environmental, etc.), to primary research (involving personal interaction with stakeholders and potential users, field observation and co-design workshops), to concepting and prototyping (requiring ideation, material and digital prototyping), to testing (with stakeholders) to finally development (of a service, product, communication or a system). The crucial question in Design Education is whether GenAI comes in to support all the activities of the design process, or it is concentrated on specific phases, and which ones in particular? Also, how does the use of GenAI affect students' perception of authorship and ownership of their projects? To address these questions a specific investigation of design students practices with GenAI tools is required.

## 2. GenAI Usage Survey in the Design School at Politecnico di Milano

In September/October 2025 we ran a survey in the School of Design of Politecnico di Milano. The Australian survey by Chung et al. [7] (available under creative commons and following previous authorisation from the authors) was adopted for comparison purposes, but modified in parts to address some domain-specific questions regarding the practices of design students and their perception of the effects of GenAI usage on creativity and authorship. In particular we asked students for which phases of a design process they used Generative AI tools (e.g. research, brainstorming, prototyping..) and whether they felt that using GenAI affected their creativity and their sense of project ownership. The questionnaire comprised 36 questions and was proposed both in English and Italian to students enrolled in the Design School. It was completed by 280 students including undergraduates and masters students from all branches of design and all years of enrolment. *The full questionnaire is available in the annex. The full set of responses are available upon request to the author.*

Reported here are a selection of the findings from the English version that was answered by 223 students. A comparison with the Italian results showed no significant differences, so we decided to report here only the English survey results. Many Italian students responded to the English version survey, the respondents therefore included both domestic and international students.

Regarding frequency of use, the majority of respondents reported engaging with GenAI tools on a daily or weekly basis (71%) for academic purposes, and regularly for personal use (59%).

Table 1. Distribution of responses to "**How often do you use GenAI for the following general reasons**" (n=223 respondents)

| Category | Daily | Weekly | Monthly | Less often | Never |
|---|---|---|---|---|---|
| University Studies | 29% | 42% | 11% | 12% | 6% |
| Work | 16% | 31% | 11% | 14% | 28% |
| Personal reasons | 29% | 30% | 12% | 16% | 12% |

The specific tasks for which GenAI is used present the usual range of information seeking, text editing, translation and synthesising. Coding, image generation and other creative media were significantly lower. Activities related to research such as finding information, summarising notes and brainstorming were all selected by more than two thirds of respondents, writing activities such as editing text, translating or producing written answers were selected by more than one in two of the students. While, surprisingly, tasks involving the generation of images, audio, music, video, games or code, were unexpectedly selected by less than one in three respondents.

Table 2. Distribution of responses to "**Select how you have used GenAI for doing the following tasks in your studies** (*tick all that apply*)" (n= 231 respondents)

| Response Option | Count | % of Respondents |
|---|---|---|
| Editing or improving writing | 179 | 77.5% |
| Finding information or conducting research | 172 | 74.5% |
| Generating ideas and examples (brainstorming) | 157 | 68.0% |
| Teaching me how to do something (step-by-step) | 150 | 64.9% |
| Summarising notes, readings or other materials | 141 | 61.0% |
| Generating written answers or content for studies | 122 | 52.8% |
| Translating text or audio into another language | 117 | 50.6% |
| Generating/error-checking code, formulas, calculations | 88 | 38.1% |
| Generating/editing artwork, images, diagrams, slides | 76 | 32.9% |
| Transcribing lectures or class discussions | 73 | 31.6% |
| Completing part or all of an assignment | 43 | 18.6% |
| Generating/editing audio, music, video, creative media | 29 | 12.6% |
| Creating games, software or apps | 17 | 7.4% |
| Other (specified a different activity) | 9 | 3.9% |

ChatGPT emerged as the near-universally adopted tool; however, respondents also reported using a broad range of other systems, including Claude, Perplexity, and Microsoft Copilot, as well as GenAI functionalities embedded in third-party applications such as Figma, Photoshop and NotebookLLM. The primary use cases centred on text generation and editing, and information retrieval, which included the summarisation of readings and the sourcing of literature and reports. Specialised tools for multimedia generation are mentioned by very few respondents confirming the previous answers on the activities done with the support of GenAI tools.

Table 3: Distribution of responses to the question " **What GenAI tools are you using** (multiple choices possible) (n=228 respondents, 513 total tool mentions, 7 non-users)

| Tool | Count |
|---|---|
| ChatGPT | 206 |
| Figma | 65 |
| NotebookLM | 49 |
| Perplexity | 40 |
| Microsoft Copilot | 40 |
| Midjourney | 38 |
| Claude | 38 |
| Gemini | 17 |
| Lovable | 4 |
| Photoshop AI | 3 |
| DeepSeek | 3 |
| Google AI Studio | 2 |
| Sora | 2 |
| Notion | 1 |
| Stable Diffusion | 1 |
| Runway | 1 |
| Adobe Firefly | 1 |

Interestingly, (and similarly to the Australian results) respondents displayed a strong sense of confidence in their ability to use these tools, while simultaneously expressing considerable distrust of AI-generated outputs. This distrust is also reflected in their reported behaviour of systematically editing and verifying the results produced by these systems.

At the survey questions asking wether they feel that they are capable of obtaining the outcomes they need, the majority felt they were able to do it. However the level of trust in the outputs is very low as 65% of the responsdents indicate that they do not trust the outputs. And 85% of respondents indicate that they systematically modify what GenAI produces as output to suit their needs.

Table 4: Distribution of responses to the question " **When I use GenAI I feel that**" (n=202 respondents):

| | Strongly disagree | Somewhat disagree | Neither agree nor disagree | Somewhat Agree | Strongly agree |
|---|---|---|---|---|---|
| I am successful in making GenAI create outcomes that meet my needs | 9% | 14% | 24% | 41% | 12% |
| I am successful in making GenAI modify and improve its outcomes to meet my needs | 11% | 10% | 17% | 49% | 13% |
| I trust what GenAI generates (i.e., that it is accurate, reliable, appropriate) | 27% | 36% | 20% | 16% | 2% |

Table 5: Distribution of responses to the question " **I change or or edit what GenAI produces to suit my needs**" (n=202 respondents):

| Response | % |
|---|---|
| Never | 4% |
| Sometimes | 4% |
| Half of the time | 6% |
| Most of the time | 42% |
| Always | 44% |

Mistrust in the results provided by GenAI tools is also confirmed by the reasons that keep respondents from using them. The fear of inaccuracy is the main reason chosen when asked what discourages students from using GenAI for their studies. 70% of the respondents are highly concerned about inaccuracy. This is an important result as it shows a critical perspective on the outputs of GenAI and can shed light on further responses concerning the perception of maintaining ownership of projects.

Table 6: Distribution of responses to the question " **I am discouraged from using GenAI in my studies because**" (n=219 responses)

| Reason | Not at all | A little | A moderate amount | A lot | A great deal |
|---|---|---|---|---|---|
| **It can be inaccurate or make things up** | 2.7% | 9.9% | 16.7% | 36.9% | 33.8% |
| **I want to do it myself** | 7.7% | 14.0% | 25.7% | 26.1% | 26.6% |
| **I have ethical concerns** | 16.2% | 19.8% | 17.6% | 17.1% | 29.3% |
| **I am worried about data privacy** | 20.3% | 25.7% | 17.6% | 15.3% | 21.2% |
| **It can't do what I need it to do** | 14.0% | 23.4% | 31.5% | 20.3% | 10.8% |
| **I can't afford access** | 26.6% | 24.8% | 22.5% | 14.0% | 12.2% |
| **I don't have access** | 36.9% | 24.8% | 16.2% | 12.2% | 9.9% |
| **I am worried about breaking rules** | 36.0% | 27.9% | 14.0% | 13.5% | 8.6% |
| **I don’t know what tools to use** | 34.7% | 26.6% | 19.8% | 14.4% | 4.5% |
| **I don’t know how to use it confidently** | 38.3% | 24.8% | 21.2% | 11.3% | 4.5% |
| **Other team members disapprove** | 66.7% | 15.5% | 12.3% | 3.2% | 2.3% |

When asked about their use of GenAI for specific **Design Tasks**, the findings corroborate the patterns identified in the literature reviewed above. Respondents demonstrated a clear preference for GenAI-assisted research and writing tasks over other activities such as coding or, surprisingly, image generation.

Table 7: Distribution of responses to the question
"**Select how you have used GenAI for doing the following specific Design activities** *(tick all that apply)*"
(n=221 respondents, 1076 unique answers)

| **Answer Choice** | **Count (Frequency N=1076)** | **% of Respondents (N = 221)** |
|---|---|---|
| Searching for articles, papers and case studies | 153 | 69.2% |
| Generating ideas and examples (e.g. brainstorming) | 150 | 67.9% |
| Summarising and analysing reports, articles, books | 144 | 65.2% |
| Understanding a topic | 142 | 64.3% |
| Correcting my documents | 106 | 48.0% |
| Organising my primary research (recruitment, interview guides, preparing surveys) | 74 | 33.5% |
| Generating images, diagrams or technical drawings | 70 | 31.7% |

| Transcribing, analysing and summarising user, market, or trend research | 60 | 27.1% |
|---|---|---|
| Organise groupwork (plan activities, manage the project) | 40 | 18.1% |
| Generating user journeys, workflows, scenarios | 39 | 17.6% |
| Rapid prototyping | 27 | 12.2% |
| Development and implementation | 25 | 11.3% |
| Creating a presentation | 20 | 9.0% |
| Testing a prototype | 13 | 5.9% |
| None of the above | 2 | 0.9% |
| Generating code | 1 | 0.5% |

The use of Generative AI assistants for research is not only significant in the frequency in which it is employed, it is also perceived as transforming the way research is being conducted. 72% of students say that they consider that research has changed since using GenAI. Brainstorming is also mentioned frequently (73%) as a supported task. Brainstorming is usually linked to the initial phase of research as it allows designers to explore the field of possibilities in their interventions..

Table 8: Distribution of responses to "**I feel the way I do research has changed since I started using GenAI**" (n= 219 responses)

| Response | % |
|---|---|
| Strongly disagree | 5% |
| Somewhat disagree | 9.6% |
| Neither agree nor disagree | 13.2% |
| Somewhat agree | 38.8% |
| Strongly agree | 33.4% |

In the same vein, GenAI is also seen as a useful tool for clarifying concepts and helping to better understand topics. 80% of the respondents agree that they find it easier to understand some topics since they started using GenAI tools.

Table 9: Distribution of responses to "**I feel that the way I brainstorm ideas, concepts and generate mockups has changed since I use GenAI**" (n=223)

| Response | Percentage |
|---|---|
| Strongly agree | 40% |
| Somewhat agree | 33% |
| Neither agree nor disagree | 17% |
| Somewhat disagree | 7% |
| Strongly disagree | 3% |

The following questions address a very crucial aspect of the design process and work and the effect that the use of GenAI has on students' perception of their **creativity** when they engage with GenAI tools. The perception of individual or collective (when working in groups) **ownership** of projects is also a factor that could be affected by the use of AI.

With respect to questions of ownership and creativity, respondents' opinions were divided; however, only 31% reported a perceived loss of ownership or creative agency when using GenAI tools. This is a very significant result for Design Education. These responses should be put in relation with the data regarding the GenAI activities less frequently selected, namely creating images, developing and testing prototypes or generating user journeys and scenarios. Furthermore, even in tasks often supported by AI, such as information seeking and writing, students say they systematically correct, edit and improve the GenAI output. Also, as seen below, they say that they use the tools to start them off or for repetitive task. This seems to suggest that there are deliberate decisions on the stages of the design process that students keep for themselves as an integral part of their creative process. The fact of systematically validating and augmenting outputs also suggests that students want to retain control, thus reducing the sense of loss of ownership of their projects.

Table 10: Distribution of responses to the question"**I feel that I am less creative since I started using AI**." (n= 219)

| Response | Percentage |
|---|---|
| Strongly agree | 6.4% |
| Somewhat agree | 24.7% |
| Neither agree nor disagree | 28.8% |
| Somewhat disagree | 22.4% |
| Strongly disagree | 17.8% |

Table 11: Distribution of responses to the question""**I feel that I have less ownership of my projects since I started using AI**." (n= 223)

| Response | Count | Percentage |
|---|---|---|
| Strongly agree | 21 | 9% |
| Somewhat agree | 52 | 22.3% |
| Neither agree nor disagree | 58 | 24.9% |
| Somewhat disagree | 51 | 21.9% |
| Strongly disagree | 51 | 21.9% |

Table 12: Distribution of responses to the question""**I feel that I retain ownership of my projects when I use GenAI to**:" (n= 218)

| | Strongly Agree | Somewhat Agree | Neither Agree nor Disagree | Somewhat Disagree | Strongly Disagree | Not applicabl |
|---|---|---|---|---|---|---|
| **Only to correct my work** | 25.8% | 35.5% | 15.2% | 7.8% | 7.4% | 8.3% |
| **Check I have not missed anything** | 29.8% | 41.0% | 10.2% | 7.0% | 5.1% | 7.0% |
| **Only to start me off** | 14.4% | 41.4% | 20.5% | 10.2% | 6.0% | 7.4% |
| **Only for repetitive tasks** | 22.5% | 38.1% | 17.4% | 6.4% | 6.0% | 9.6% |
| **For tasks I would not know how to do** | 13.0% | 36.6% | 23.1% | 13.0% | 8.8% | 5.6% |
| **For tasks I don't care about** | 11.6% | 31.2% | 26.0% | 9.8% | 10.7% | 10.7% |
| **I don't feel I retain ownership** | 4.2% | 12.7% | 19.3% | 15.6% | 23.6% | 24.5% |

Examining the primary motivations for using Generative AI tools, time-saving and ease emerged as the dominant factors. Specifically, 63% of respondents indicated that speed was a key driver for using generative AI in their studies, while over 50% selected the answer "makes things easier". In contrast, only 34% reported using GenAI because it "makes things easier" or to get unstuck. Less than 30% use are motivated to use GenAI to accomplish tasks they would not be able to complete independently, to improve grades or improve the quality of their work. These findings suggest that GenAI is not perceived as a substitute for students' own capabilities, but rather as a tool that enhances the speed with which they execute tasks they could potentially carry out autonomously.

Table 13: Distribution of responses to the question""**I am motivated to use GenAI in my studies because it:**" (n= 218)

| Reason | Not at all | A little | A moderate amount | A lot | A great deal |
|---|---|---|---|---|---|
| Makes things easier | 6% | 22% | 38% | 23% | 11% |
| Makes things faster | 5% | 14% | 18% | 38% | 25% |
| Improves the quality of my work | 28% | 28% | 27% | 11% | 6% |
| Helps me get unstuck / overcome blocks | 8% | 26% | 33% | 23% | 11% |
| Helps me do things I could not do otherwise | 28% | 20% | 24% | 20% | 8% |
| Helps me overcome accessibility challenges | 32% | 23% | 24% | 16% | 5% |
| Improves my English language skills | 47% | 16% | 17% | 14% | 6% |
| Improves my grades | 51% | 26% | 15% | 6% | 3% |
| Reduces my stress | 29% | 29% | 21% | 12% | 8% |
| Is enjoyable | 29% | 26% | 25% | 14% | 7% |
| Helps me feel more confident | 37% | 36% | 13% | 7% | 7% |

## 3. Analysis of the survey results

The statistical analysis of the survey results did not detect any significant difference in the patterns of use and attitudes towards Generative AI tools, by gender, year of enrolment, design domain or language. The results align with the findings of numerous studies on generative AI in education, namely the frequent and systematic use of GenAI for academic purposes, with a predominant focus on information seeking, topic understanding, and writing tasks. A convergence on few well known tools: ChatGPT, Claude, Gemini and limited use of coding tools and services also show a focus on research activities.

Regarding the questions which were drafted to be more specific to design education, there are some interesting findings that have important implications for how the processes of design may have already started to change as a consequence of the adoption of these tools.

1. Design students are **primarily using GenAI for research tasks**, writing and brainstorming and less so for other phases of the design process,
2. The **perception of ownership and creativity are relatively unchanged** even for frequent users of GenAI.

A correlation analysis was conducted examining responses related to ownership and creativity alongside reported AI-assisted activities. As anticipated, a strong association was found between students' perceived changes in research practices and activities such as finding and synthesising articles, organising primary and secondary research, writing, and identifying case studies. None of these activities, however, were significantly correlated with a sense of loss of creativity or ownership over one's work. Using GenAI for brainstorming was not correlated with changed perception of creativity and ownership. The only activity significantly associated with such feelings was the use of generative AI for prototype generation.

At the question regarding which activities they used GenAI for, 75% of the respondents selected: 1. Searching for articles, papers and case studies, 2. Summarising and analysing reports, articles, books, 3. Correcting my documents, 4. Organising my primary research (recruitment, interview guides, preparing surveys), 5. Generating ideas and examples (e.g. brainstorming) as design activities they do with GenAI. The same number considers GenAI tools useful for saving time and effort. However, the search for speed and efficiency has no effect on the perception of ownership and creativity.

For design studies this raises major questions around the role of the research phases in the design process. Research has been a hallmark of user centred design as it positions the creative phase in a context that has to be well understood and analysed in order to situate a project in terms of significance, relevance and impact. Research is also a significant element of the educational approach to design of the School of Design of Politecnico di Milano which puts a strong emphasis on "metadesign" in all curricula. The survey results suggest that students may not perceive the research phase as an integral part of their creative process and therefore it constitutes an activity that *can be or needs to be* supported by GenAI tools.

We can give two interpretations to this result: 1) research, especially secondary, is currently considered too complex and time consuming and therefore students feel that they require the support of AI assistants, or 2) the research phase is not sufficiently valued as a design activity and therefore can be offloaded in part to Generative AI assistants. The comments that accompany the question regarding the perceived change when doing research all focus on the speed with which literature is accessed and research completed. In evaluating this result we should also take into account the huge increase in available information, articles, reports that has characterised the last years and the difficulty in navigating and analysing this mass of material. However, the need to accelerate the process is also potentially an indication of a decreasing appreciation of the importance of in-depth knowledge of the context of a design project.

In both cases we are confronted with a major educational question as the reliance on GenAI tools for secondary research must be better acknowledged and integrated in the learning processes. It also raises the issue of the role of primary research in the design process, which as an activity that remains less affected by GenAI practices, could be strengthened.

A focused analysis of the use of GenAI during research activities, shed some further light on this issue.

## 4. Comparing the survey results to the use of GenAI in design projects

During the Research Methods Course of the Masters in Product Service System Design (PSSD) of Politecnico di Milano, of the academic year 2025/2026, we asked students to keep a journal of their GenAI usage in carrying out their coursework and deliverables. The course requires students to produce two main assignments carried out in groups of 5. The first is an in depth secondary research on a specific question that is the topic of a a product and service project they will develop in a related course called the Innovation Studio, presented as a 20 page report. The second is a primary research consisting of interviews, in situ visits, and observations, to enrich their understanding of the social and physical context in which their project for the Innovation Studio is set (presented as a report and a multimedia production). The overall topic of the research and studio that year was "Water scarcity in Sicily". In their secondary research groups were assigned specific topics to investigate such as the water infrastructure of Sicily, the hydrological conditions of the island, the emergency policies developed in the last 5 years to address episodes of severe drought, EU legislation on the water crisis, etc. While the topics of the secondary research were assigned by the lecturer, the scope and object of the primary research was chosen by the groups as a function of their project. Groups therefore interviewed farmers, local administrations, utilities and visited orchards, farms and Sicilian towns. For each assignment students had 3 to 4 weeks to complete their work, results were provided as written reports, presentations and multimedia content (podcasts, videos, magazines, etc.).

Two forms had been prepared with questions on the GenAI tools used, verification processes and their added work on the output of GenAI (see Annex 2 and 3). More importantly the forms (designed by the author) provided a division in 15 subtasks that a secondary and a primary research require such as finding references, summarising papers, analysing data from literature, creating an outline, or creating an interview protocol, transcribing interviews, analysing observations and videos, writing reports, etc.

For each assignment some groups provided a joint form illustrating their collective use of GenAI and some preferred to produce individual forms. Overall 100 students completed the form which was mandatory although it was made clear that the content of the responses would not be considered in the evaluation and did not contribute to their final marks.

## 4.1 Analysis of the journals

The answers to the journals have to be obviously taken with some caution as they are self-reports in a context of evaluation and therefore have an element of self-validation and are an expression of what may be considered an acceptable use of GenAI for an assignment. In class numerous GenAI based research tools had been presented and experimenting with GenAI was part of the curriculum. While students were made aware of the rules regarding plagiarism, and self-disclosure when using GenAI tools for their assignments, the course conveyed the idea that experimenting with GenAI was acceptable as long as it was reported.

The journals were filled very unequally, some were highly detailed and other quite generic. Thus, as a corpus, they provide a only partial insight into current practices. However they do offer some indications on the phases in which GenAI is used and is deemed socially and academically acceptable.

If we look at the patterns of GenAI use reported in the forms, we see common strategies emerging:

**GenAI for Identifying literature and Information Gathering**

Students use GenAI tools to explore the topics and quickly identify relevant sources in the literature be it academic, administrative or reports from associations and NGOs.

ChatGPT, Gemini, Deepseek, Perplexity, and Elicit are the most popular tools for:

- Brainstorming research questions and themes.
- Generating initial lists of sources, reports and papers.
- Finding hard-to-locate documents (e.g., NGO reports, administrative and legal documents).

The common strategy in this phase seems to be:

- Start with GenAI to get a broad overview or initial list of sources.
- Cross-check GenAI suggestions with academic databases (Google Scholar, JSTOR) or institutional reports.
- Use multiple GenAI tools (e.g., ChatGPT + Perplexity) to diversify insights and reduce bias.

**AI for Summarising**

Students rely on GenAI to condense lengthy papers and extract the key points.

ChatGPT, Scispace, and NotebookLM are used to:
- Summarise academic papers and reports.
- Highlight main arguments, methodologies, and results.
- Create comparison tables or affinity maps for synthesising findings.

The common strategy in this phase seems to be:

- AI-generated summaries are reviewed for accuracy and correctness.
- Students add their own notes or delete AI-generated content that doesn't align with their research focus.

**AI for Writing and Editing**

GenAI tools are widely used to refine and correct drafts.

ChatGPT, DeepL Write, Quillbot, and Grammarly help with:
- Restructuring arguments for coherence.
- Paraphrasing or rephrasing sentences.
- Checking grammar and correcting language.
- Translating non-English sources (e.g., Italian articles into English), and translating own text into English.

The common strategy in this phase seems to be:

- Students say that they write the first draft independently and then use GenAI for editing and enhancement.
- AI outputs are reviewed to ensure they don't misrepresent the original ideas.

**AI for Visualisation**

Students use GenAI to create maps, diagrams, and data visualisations.

Canva AI, Google My Maps, and Figma are used to:
- Generate maps (e.g., locations of desalination plants in Sicily).
- Create diagrams or affinity maps for data organisation.
- Design presentations (e.g., Canva AI for slide templates).

The common strategy in this phase seems to be:

- AI tools generate raw data or coordinates (e.g., ChatGPT for CSV files).
- Students customise visuals to fit their research narrative.

**AI for Technical Understanding**

Students use GenAI to decode complex terminology and grasp technical concepts:

ChatGPT and other GenAI tools help with:

- Explaining technical terms in academic papers.
- Summarising technical reports (e.g., water management strategies).

**AI for Transcribing and Analysing interviews**

Students use GenAI to transcribe recordings of interviews and for analysing and comparing transcriptions:

ChatGPT and other GenAI tools help with:

- Transcribing into text the audio recordings of the interviews.
- Summarising the interviews.
- Comparing the outputs of the interviews

The common strategy in this phase seems to be:

- use GenAI for automatic transcription and then check output with audio recording

- use GenAI to synthesise transcripts and then check and edit

- Use GenAI to compare transcripts and then check and edit

In summary, GenAI tools are extensively used to identify and analyse the literature, transcribing interviews and for correcting and improving texts. Maps, diagrams and visualisations are also supported by GenAI tools.

A general characteristic of the workflows described by the students is the emphasis on verification and collaboration. While GenAI tools are employed to generate initial literature lists, summarise papers, transcribe interviews, and refine language, students say that they consistently cross-check GenAI outputs for accuracy, credibility, and relevance. All respondents said that they manually verified AI-generated sources or summaries, often through group discussions or by consulting institutional reports and academic databases. Furthermore, collaboration remains a cornerstone of the research process, with students using shared platforms like Google Docs, Miro, and Figma to compile findings, avoid repetition and refine arguments collectively.

Notably, not all research stages benefit equally from GenAI integration. While phases like *finding initial information*, *summarising papers*, *transcribing and summarising interviews,* and *editing/polishing* show near-universal GenAI adoption, others, such as *developing an argument,* remain more often in control of the students. Similarly in most cases the interpretation of the interviews and observations during primary research is done without GenAI assistance as is the structure of the data collection.

What we see reported by students, is a heavy reliance on GenAI agents during secondary research for finding and organising information and greater autonomy in the tasks that require planning and interpretation. This is in line with the responses in the survey that indicate that efficiency and ease are the main drivers. It would tend to suggest that students are using GenAI to quickly scan the literature significantly reducing the time to find and summarise relevant sources. Writing is also significantly supported by GenAI tools, in editing and correcting but also in enhancing and rewriting drafts. On the other hand, primary research is much more traditional. Students did not rely on the numerous GenAI tools that are available to generate synthetic participants, create personas or generate videos, (tools that had been presented in class). Primary research activities were mostly supported by GenAI in the transcription and synthesis of audio transcripts, time consuming tasks that fit well the drive for time saving reported in the survey. However, planning, execution and analysis of primary research are mostly conducted with limited GenAI support.

## 5. Conclusions

The results from the Survey done at the School of Design of the Politecnico di Milano have highlighted the role of Generative GenAI assistants especially in the research phases of students' design process. Speed, efficiency and ease are the main drivers for using the GenAI tools in finding relevant literature, summarising papers, and supporting the production of texts. Assistance by GenAI tools in these tasks however **was not perceived as affecting the sense of ownership of a project or creativity**. When integrating the results from self-reporting of activities during research tasks, we have found confirmation that GenAI tools are used extensively for these tasks. In addition transcribing and analysing audios and creating visualisations, maps and diagrams are also quite systematically supported by GenAI tools.

Both in the anonymous survey and in the self-reporting forms, students say that they systematically verify the outputs of the GenAI systems and improve or work on results provided. This can be done individually or collaboratively as a group. Therefore references are checked and texts rewritten or critically examined. The reliability of these claims can be supported by answers about the limited level of trust that students bestow on the tools they use.

### 5.1 Open questions

These findings raise a number of questions on how, as design educators, we should integrate these new practices in teaching the design process.

1. One of the preoccupations expressed by many educators, and by the members of the Design School at Politecnico di Milano lecturers (in a survey run in 2025) is that students should acquire a critical view of AI. In a sense, these findings should be reassuring, as students report that they systematically verify and "augment" the outputs produced by GenAI systems. Group discussions and analysis also seem to play an important role in validating the contribution of GenAI tools. The issue is how do we ensure that these practices continue and, as time goes by, and a greater familiarity with the tools is acquired, students don't fall into complacency and start accepting GenAI outputs at face value without evaluating them? We also need to continue in a longitudinal approach to administer the survey, to monitor the evolution of students' practices and attitudes to GenAI in their design education. This means administering the survey on a yearly basis and evolve it to take into account the constant transformations of AI tools and services.

2. Another often raised issue is that of delegating tasks and effort thus limiting the learning process to assembling content found and generated by others rather than working through the difficulties of finding sources, reading and synthesising autonomously. The secondary research reports produced in the Methods Course which were developed with significant assistance of GenAI were extremely good and at a very high standard. But the question remains with regards to the transformation of this content into operational and actionable knowledge. Are students in fact learning less because part of the work is done for them? And in that case how do we ensure that the support they receive from GenAI tools can be leveraged for better learning?

3. The most challenging issue for design education, is the interaction between the support from GenAI tools and creativity or project ownership. This requires serious consideration. As educators we may need to re-examine the way we teach the relationship between research, concepting and idea generation. If we believe that research is an integral part of the creative process, providing not only the justification but the spark for innovative creation, we may need to understand how to put the emphasis on the translation of information into insights and into ideas. This may allow students to better value the time and effort that the secondary research phase requires.

4. Primary research was less affected by the deployment of GenAI compared to secondary research for obvious reasons. Planning the research, interacting with people, doing observations in real contexts and interpreting findings, are practical tasks that cannot be easily automated. Should we consider strengthening this part of the research to enable students to be more directly involved in the collection of their own data? Would this personal and human experience allow them to attribute more value to the research activities they need to carry out to build their understanding of the context of their projects?

5. Another avenue to address the issue of creativity and ownership, regards the choice of GenAI tools themselves. Students are using the main big platforms, ChatGPT above all, with very little adaptation and invention. There is a case to be made for ensuring that students become less of "consumers" of GenAI tools and more "makers" of their own tools. There are increasingly possibilities to customise and create personal and specific GenAI based apps to execute tasks that are tailored to individual or group needs. From low or no code app creation, to creating custom gpts, to using and adapting open source models, there are numerous ways in which design students could customise GenAI tools to their needs. This would bring their interaction with GenAI tools to another level in which part of their learning would be to create tools for their projects. How can we encourage students to experiment with developing their own GenAI tools?

6. Finally we need to have more insights on how students are using GenAI for brainstorming and other phases of the design process in order to evaluate the transformations across all the creative cycle. We would encourage other design courses to experiment with different approaches to collect data on students' practices with AI.

**ANNEXES**

Annex 1: Full Survey at Politecnico di Milano Design School in September October 2025 : Questions

# Generative AI in Design Higher Education

## Introduction

This survey was created as part of a multi-institutional research project: "Student Perspectives on AI in Higher Education". The project is a collaborative effort of some major European Design Schools: Politecnico di Milano, TU Delft, Aalto University. The project aims to understand higher education students' relationships with Generative Artificial Intelligence (GenAI).

The problem at the heart of this research is the rapid integration of GenAI technologies in educational settings, which out-paces our understanding of how students are navigating these changes. Many existing discussions of GenAI in education rely on assumptions or simplistic views of human-technology relationships. Our project investigates higher education students' views, beliefs, and uses of GenAI guided by an appreciation of the complexity and diversity of student experiences and en-gagement with GenAI. In particular we would like to understand how design students are integrating this technology in their projects and creative processes and the opinions they have on how this may be transforming their education.

## Consent

You are invited to take part in the collaborative research project exploring university students' experiences with Generative Artificial Intelligence (GenAI) in education.
All the data collected is anonymous. Please review the Informed Consent document prior to participating in this research study. The survey takes 15mns to be completed, Once you have started the survey you can stop and come back to complete the survey later.

1. I have read the **Informed Consent** and: *

- ◯ Yes, I wish to take part in this study and complete the survey
- ◯ No, I do not wish to take part in this study

## Demographics

Your responses are anonymous and there is no collection of the IP address

2. What is your course level?

- ◯ Undergraduate

- ◯ Masters
- ◯ Doctorate

3. Which of the below most closely matches your main area of study?

- ◯ Product Design
- ◯ Communication
- ◯ Fashion and textile
- ◯ Interaction Design
- ◯ Service Design
- ◯ Inerior Design
- ◯ Visual Arts
- ◯ Media and Design
- ◯ Business and Design
- ◯ Other

4. How many years have you been enrolled in your degree? (full-time equivalent)

- ◯ I have just started my degree
- ◯ 1 year
- ◯ 2 years
- ◯ 3 years
- ◯ 4 years or more

5. What is your enrolment type?

- ◯ Domestic student
- ◯ International student

6. How do you describe your gender identity?

- ◯ Woman
- ◯ Man
- ◯ Non-binary
- ◯ Gender diverse
- ◯ Prefer not to say

## How you use Gen AI

**This survey is about Generative Artificial Intelligence.** Generative Artificial Intelligence (GenAI) is a branch of AI that is usu-ally used to create new data or content. Humans can use GenAI to create a wide range of outputs, including text, images, au-dio, videos, computer code, and analysis of writing and data. Examples include chatbots like ChatGPT, Microsoft Copilot, or image creation tools like Midjourney that create personalised responses to user prompts.
**GenAI literacy** The questions on this page ask about your experiences with using GenAI. When answering these questions, think about the GenAI tools you have used and your experience overall.

7. I know how to use GenAI to:

| | Strongly agree | Somewhat Agree | Neither agree nor disagree | Somewhat disagree | Strongly disagree |
|---|---|---|---|---|---|
| Answer my questions | ◯ | ◯ | ◯ | ◯ | ◯ |
| Create written text I can use | ◯ | ◯ | ◯ | ◯ | ◯ |
| Create computer code or other technical outputs | ◯ | ◯ | ◯ | ◯ | ◯ |
| Analyse documents or data | ◯ | ◯ | ◯ | ◯ | ◯ |
| Create images or other visual media | ◯ | ◯ | ◯ | ◯ | ◯ |

8. When I use GenAI I feel that

| | Strongly agree | Somewhat Agree | Neither agree nor disagree | Somewhat disagree | Strongly disagree |
|---|---|---|---|---|---|
| I am **successful** in making Gen AI **create** outco mes that meet my needs | ◯ | ◯ | ◯ | ◯ | ◯ |
| I am **successful** in making Gen AI **modify** and i mprove its outc omes to meet my needs | ◯ | ◯ | ◯ | ◯ | ◯ |
| I **trust** what GenAI generates (i.e., that it is accurate, reliable, appropriate) | ◯ | ◯ | ◯ | ◯ | ◯ |

9. I **change or edit** what GenAI produces to suit my needs

- ◯ Always
- ◯ Most of the time
- ◯ Half of the time
- ◯ Sometimes
- ◯ Never

10. What Gen AI tools are you using (multiple choices possible)

- ☐ ChatGPT
- ☐ Claude
- ☐ Perplexity
- ☐ Midjourney
- ☐ Microsoft Copilot

- [ ] Lovable
- [ ] Figma
- [ ] NotebookLLM
- [ ] Other

11. Could you list some other tools you use

12. How often do you use GenAI tools for the following general reasons:

| | Daily | Weekly | Monthly | Less often | Never |
|---|---|---|---|---|---|
| University Studies | ○ | ○ | ○ | ○ | ○ |
| Work | ○ | ○ | ○ | ○ | ○ |
| Personal reasons | ○ | ○ | ○ | ○ | ○ |

13. Select how you have used GenAI for doing the following tasks in your studies *(tick all that apply)* *

- [ ] Editing or improving writing
- [ ] Generating written answers or content for studies
- [ ] Generating or error-checking code, formula, mathematical calculations or data analysis
- [ ] Generating or editing artwork, images, diagrams, graphs or slides
- [ ] Generating or editing audio, music, video or other creative media
- [ ] Creating games, software or apps
- [ ] Generating ideas and examples (e.g. brainstorming)
- [ ] Finding information or conducting research
- [ ] Transcribing lectures, class discussions or other content
- [ ] Translating text or audio into another language
- [ ] Summarising notes, readings or other materials
- [ ] Teaching me how to do something, for example step-by-step instructions
- [ ] Completing part or all of an assignment
- [ ] Other

14. How **helpful** was GenAI for doing these **tasks** in your studies

| | Very helpful | Somewhat helpful | Neither helpful nor unhelpful | Somewhat unhelpful | Very unhelpful | Does not apply |
|---|---|---|---|---|---|---|
| Editing or improving writing | ◯ | ◯ | ◯ | ◯ | ◯ | ◯ |
| Generating written answers or content | ◯ | ◯ | ◯ | ◯ | ◯ | ◯ |
| Generating or error-checking code, formula, mathematical calculations or data analysis | ◯ | ◯ | ◯ | ◯ | ◯ | ◯ |
| Generating or editing artwork, images, diagrams, graphs or slides | ◯ | ◯ | ◯ | ◯ | ◯ | ◯ |
| Generating or editing audio, music, video or other creative media | ◯ | ◯ | ◯ | ◯ | ◯ | ◯ |
| Creating games, software or apps | ◯ | ◯ | ◯ | ◯ | ◯ | ◯ |
| Generating ideas and examples (e.g. brainstorming) | ◯ | ◯ | ◯ | ◯ | ◯ | ◯ |
| Finding information or conducting research | ◯ | ◯ | ◯ | ◯ | ◯ | ◯ |
| Transcribing lectures, class discussions or other content | ◯ | ◯ | ◯ | ◯ | ◯ | ◯ |
| Translating text or audio into another language | ◯ | ◯ | ◯ | ◯ | ◯ | ◯ |
| Summarising notes, readings or other materials | ◯ | ◯ | ◯ | ◯ | ◯ | ◯ |
| Teaching me how to do something, for example step-by-step instructions | ◯ | ◯ | ◯ | ◯ | ◯ | ◯ |
| Completing part or all of an assignment | ◯ | ◯ | ◯ | ◯ | ◯ | ◯ |
| Other | ◯ | ◯ | ◯ | ◯ | ◯ | ◯ |

15. Select how you have used GenAI for doing the following **specific Design activities** *(tick all that apply)* *

- [ ] Understanding a topic
- [ ] Searching for articles, papers and case studies
- [ ] Summarising and analysing reports, articles, books
- [ ] Correcting my documents
- [ ] Organising my primary research (recruitment, interview guides, preparing surveys)
- [ ] Transcribing, analysing and summarising user, market, or trend research
- [ ] Generating ideas and examples (e.g. brainstorming)
- [ ] Generating images, diagrams or technical drawings
- [ ] Generating user journeys, workflows, scenarios
- [ ] Creating a presentation
- [ ] Rapid prototyping
- [ ] Testing a prototype
- [ ] Development and implementation
- [ ] Organise groupwork (plan activities, manage the project)
- [ ] Other

16. I am **motivated** to use GenAI in my studies because it *

| | Not at all | A little | A moderate amount | A lot | A great deal |
|---|---|---|---|---|---|
| Makes things easier | ◯ | ◯ | ◯ | ◯ | ◯ |
| Makes things faster | ◯ | ◯ | ◯ | ◯ | ◯ |
| Improves the quality of my work | ◯ | ◯ | ◯ | ◯ | ◯ |
| Helps me get unstuck / overcome blocks | ◯ | ◯ | ◯ | ◯ | ◯ |
| Helps me do things I could not do otherwise | ◯ | ◯ | ◯ | ◯ | ◯ |
| Helps me overcome accessibility challenges | ◯ | ◯ | ◯ | ◯ | ◯ |
| Improves my English language skills | ◯ | ◯ | ◯ | ◯ | ◯ |
| Improves my grades | ◯ | ◯ | ◯ | ◯ | ◯ |
| Reduces my stress | ◯ | ◯ | ◯ | ◯ | ◯ |
| Is enjoyable | ◯ | ◯ | ◯ | ◯ | ◯ |
| Helps me feel more confident | ◯ | ◯ | ◯ | ◯ | ◯ |

17. List any other motivations *(this question is optional)*

18. I am **discouraged** from using GenAI in my studies because *

| | Not at all | A little | A moderate amount | A lot | A great deal |
|---|---|---|---|---|---|
| I am worried about breaking my university's rules | ◯ | ◯ | ◯ | ◯ | ◯ |
| It can be inaccurate or make things up | ◯ | ◯ | ◯ | ◯ | ◯ |
| It can't do what I need it to do | ◯ | ◯ | ◯ | ◯ | ◯ |
| I don't know how to use it confidently | ◯ | ◯ | ◯ | ◯ | ◯ |
| I don't know what tools to use | ◯ | ◯ | ◯ | ◯ | ◯ |
| I can't afford access to good GenAI tools | ◯ | ◯ | ◯ | ◯ | ◯ |
| I don't have access to good GenAI tools | ◯ | ◯ | ◯ | ◯ | ◯ |
| I want to do it myself | ◯ | ◯ | ◯ | ◯ | ◯ |
| I am worried about the privacy of my own data | ◯ | ◯ | ◯ | ◯ | ◯ |
| I have ethical concerns | ◯ | ◯ | ◯ | ◯ | ◯ |
| Other team members disapprove | ◯ | ◯ | ◯ | ◯ | ◯ |

19. List any other discouraging reasons *(this question is optional)*

## Role of university

These questions will help teachers, universities, and others understand if universities are supporting the acquisition of the skills to use GenAI.

20. My university provides me access to GenAI for study and learning

- Yes
- No
- I don't know

21. How does your university support you in understanding how to use GenAI in your studies?

| | Strongly agree | Somewhat agree | Neither agree nor disagree | Somewhat disagree | Strongly disagree |
|---|---|---|---|---|---|
| My university provides me with enough guidance to use GenAI effectively in my studies | ◯ | ◯ | ◯ | ◯ | ◯ |
| My university provides me with enough guidance to use GenAI effectively in my future profession | ◯ | ◯ | ◯ | ◯ | ◯ |
| I am always provided clear instructions on how I should use GenAI for assessments | ◯ | ◯ | ◯ | ◯ | ◯ |
| I have followed a course in which I was taught how to use GenAI tools | ◯ | ◯ | ◯ | ◯ | ◯ |

22. Have you done a university assessment that required you to explain whether you used GenAI in doing the assignment?

- Yes
- No

◯ I am not sure

◯ Not applicable - I have not been allowed to use GenAI in an assignment

23. When you were asked to explain how you used GenAI, what did you do

◯ I included all details of how I used GenAI ◯ I

included some details of how I used GenAI

◯ I made something up or provided incorrect details of how I used GenAI ◯

Not applicable

24. Why did you do this? *(select all that apply)*

☐ I didn't know how

☐ It was too much effort

☐ I didn't keep a record of what I did

☐ I was worried that my explanation might look like I broke the rules

☐ I don't want to seem like I need help

☐ Not applicable

☐ Other

## Learning with Gen AI

This section asks about your perceptions and attitudes towards GenAI. (*Your responses are anonymous).*

25. It is important to be able to effectively use GenAI for:

| | Strongly agree | Somewhat agree | Neither agree nor disagree | Somewhat disagree | Strongly disagree |
|---|---|---|---|---|---|
| Your university studies | ◯ | ◯ | ◯ | ◯ | ◯ |
| Your future work/ job | ◯ | ◯ | ◯ | ◯ | ◯ |

26. Do you think GenAI: *

| | Strongly agree | Somewhat agree | Neither agree nor disagree | Somewhat disagree | Strongly disagree |
|---|---|---|---|---|---|
| Makes formal education less valuable | ◯ | ◯ | ◯ | ◯ | ◯ |
| Offers significant benefits to students | ◯ | ◯ | ◯ | ◯ | ◯ |
| Makes us less intelligent | ◯ | ◯ | ◯ | ◯ | ◯ |
| Makes education fairer | ◯ | ◯ | ◯ | ◯ | ◯ |
| Increases cheating | ◯ | ◯ | ◯ | ◯ | ◯ |
| Creates tensions in teams | ◯ | ◯ | ◯ | ◯ | ◯ |
| Will make it more difficult to find a job in my field | ◯ | ◯ | ◯ | ◯ | ◯ |

27. I feel that the way I do research has changed since I started using GenAI

| Strongly agree | Somewhat agree | Neither agree nor disagree | Somewhat disagree | Strongly disagree |
|---|---|---|---|---|
| ◯ | ◯ | ◯ | ◯ | ◯ |

28. If you think something has changed, could you explain what?

29. I feel that it is easier for me to understand some topics since I started using AI

| Strongly agree | Somewhat agree | Neither agree nor disagree | Somewhat disagree | Strongly disagree |
|---|---|---|---|---|
| ◯ | ◯ | ◯ | ◯ | ◯ |

30. I feel that the way I brainstorm ideas, concepts and generate mockups has changed since I use GenAI

| Strongly agree | Somewhat agree | Neither agree nor disagree | Somewhat disagree | Strongly disagree |
|---|---|---|---|---|
| ○ | ○ | ○ | ○ | ○ |

31. If you think something has changed, could you explain what?

32. I feel that the way I collaborate with other team members has changed since I started using AI

| Strongly agree | Somewhat agree | Neither agree nor disagree | Somewhat disagree | Strongly disagree |
|---|---|---|---|---|
| ○ | ○ | ○ | ○ | ○ |

33. If you think something has changed, could you explain what?

34. I feel that I am less creative since I started using AI

| Strongly agree | Somewhat agree | Neither agree nor disagree | Somewhat disagree | Strongly disagree |
|---|---|---|---|---|
| ○ | ○ | ○ | ○ | ○ |

35. I feel that I have less ownership of my projects since I started using AI

| Strongly agree | Somewhat agree | Neither agree nor disagree | Somewhat disagree | Strongly disagree |
|---|---|---|---|---|
| ○ | ○ | ○ | ○ | ○ |

36. I feel I retain ownership of my projects when

| | Strongly Agree | Somewhat Agree | Neither Agree nor Disagree | Somewhat Disagree | Strongly Disagree | Not applicable |
|---|---|---|---|---|---|---|
| I use GenAI only to correct my work | ○ | ○ | ○ | ○ | ○ | ○ |
| I use GenAI to check I have not missed anything | ○ | ○ | ○ | ○ | ○ | ○ |
| I use GenAI only to start me off | ○ | ○ | ○ | ○ | ○ | ○ |
| I use GenAI only for the repetitive tasks | ○ | ○ | ○ | ○ | ○ | ○ |
| I use GenAI for tasks I would not know how to do | ○ | ○ | ○ | ○ | ○ | ○ |
| I use GenAI for tasks I don't care about | ○ | ○ | ○ | ○ | ○ | ○ |
| I don't feel I retain ownership of my projects | ○ | ○ | ○ | ○ | ○ | ○ |

37. Could you give an example of a recent design activity you carried out with GenAI? *(optional question)*

38. We plan to invite some participants to join an in person discussion on the topic of AI in design education. If you're willing to participate, please contact ....... . Thank you.

Annex 2 : Journal for Secondary research

| | Tool used | Verification process | Own contributions |
|---|---|---|---|
| **Understanding the question** | | | |
| **Finding initial information and scoping** | | | |
| **Creating an outline** | | | |
| **Finding detailed information** | | | |
| **Finding sources** | | | |
| **Summarising papers** | | | |
| **Data analysis** | | | |
| **Creating an overall synthesis of the readings** | | | |
| **Developing an argument** | | | |
| **Writing** | | | |
| **Creating visuals** | | | |
| **Editing/polishing** | | | |
| **Translating** | | | |
| **Outlining presentation** | | | |
| **Designing presentation** | | | |

Annex 3 : Journal for Primary research

| | Tool used | Verification process | Modifications done |
|---|---|---|---|
| **Creating a research plan/ protocol** | | | |
| **Formulating the questions** | | | |
| **Creating observation -interview protocol** | | | |
| **Recruiting** | | | |
| **Transcribing interviews** | | | |
| **Summarising interviews** | | | |
| **Comparing interview transcripts** | | | |
| **Analysing observations** | | | |
| **Data analysis** | | | |
| **Formulating interpretation** | | | |
| **Finding other research to compare results** | | | |
| **Writing Report** | | | |
| **Creating visuals** | | | |
| **Creating multimedia deliverable** | | | |
| **Translating** | | | |
| **Outlining presentation** | | | |
| **Designing presentation** | | | |